\documentclass[11pt,a4paper]{article}
\pdfoutput=1
\usepackage{jcappub}
\usepackage{bm}
\usepackage{soul}
\usepackage{indentfirst}
\usepackage{amsmath}
\usepackage{graphicx}
\usepackage{float}
\usepackage{amssymb}
\usepackage{subfigure}
\usepackage{amssymb}
\usepackage{psfrag}
\usepackage{hyperref}
\usepackage{epstopdf}
\hypersetup{
	colorlinks=true,
	linkcolor=red,
	citecolor=blue,
}
\usepackage[utf8]{inputenc}
\hypersetup{
    colorlinks=true,
    linkcolor=red,
    citecolor=blue,
}
\def\n3{\sqrt{3}}
\def\Si{\text{Si}}
\def\Ci{\text{Ci}}
\newcommand{\ux}{\frac{ux}{\sqrt{3}}}
\newcommand{\vx}{\frac{vx}{\sqrt{3}}}
\def\t{\text}

\newcommand{\sbra}[1]{\left(#1\right)}
\newcommand{\sqbra}[1]{\left[#1\right]}
\begin{document}
\title{Constraints on primordial curvature perturbations from primordial black hole dark matter and secondary gravitational waves}
\author[a]{Yizhou Lu,}
\author[a,1]{Yungui Gong\note{Corresponding author.},}
\author[b]{Zhu Yi,}
\author[a]{Fengge Zhang}

\affiliation[a]{School of Physics, Huazhong University of Science and Technology,
Wuhan, Hubei 430074, China}
\affiliation[b]{Department of Astronomy, Beijing Normal University,
Beijing 100875, China}

\emailAdd{louischou@hust.edu.cn}
\emailAdd{yggong@hust.edu.cn}
\emailAdd{yz@bnu.edu.cn}
\emailAdd{d201880055@hust.edu.cn}

\keywords{primordial black holes, gravitational waves, induced gravitational waves }
\abstract{
Primordial black holes and secondary gravitational waves can be used to probe the small scale physics at very early time.
For secondary gravitational waves produced after the horizon reentry, we derive an analytical formula for the time integral of the source
and analytical behavior of the time dependence of the energy density of induced gravitational waves is obtained.
By proposing a piecewise power-law parametrization for the power spectrum of primordial curvature perturbations,
and fitting it to observational constraints on primordial black hole dark matter, we obtain an upper bound on the power spectrum
and discuss the detection of induced gravitational waves by future space based gravitational wave antenna.
}
\maketitle
\section{Introduction}
As a result of gravitational collapse, primordial black holes (PBHs) \cite{Hawking:1971ei,Carr:1974nx,Carr:1975qj,Khlopov:2004sc}
form in a region with its density contrast at horizon reentry
during radiation domination exceeding the threshold value.
Since the temperature and polarization measurements on cosmic microwave background anisotropy only constrain the primordial perturbations to be very small at large scales,
large perturbations at small scales that cause the formation of PBHs \cite{Frampton:2010sw,Drees:2011hb,Drees:2011yz,Orlofsky:2016vbd,
Garcia-Bellido:2017aan,Garcia-Bellido:2017mdw,
Cheng:2018qof,Cheng:2018yyr,Cai:2019jah,Clesse:2018ogk,
Passaglia:2018ixg,Wang:2016ana,Wang:2019kaf,Bartolo:2018rku,
Bartolo:2018evs,Motohashi:2017kbs} are not constrained and they may produce observable secondary gravitational waves (induced GWs)
\cite{Matarrese:1997ay,Ananda:2006af,Baumann:2007zm,Bugaev:2009zh,
Bugaev:2010bb,Alabidi:2012ex,Inomata:2016rbd,Gong:2017qlj,Cai:2018dig,Cai:2019amo,Cai:2019elf,Kohri:2018awv,Drees:2019xpp,Inomata:2019ivs,
Inomata:2018epa,Inomata:2019zqy,Espinosa:2018eve,DeLuca:2019llr}.
Therefore, both PBHs and secondary GWs can be used to probe the small scale physics at very early time \cite{Mollerach:2003nq,Saito:2008jc,Saito:2009jt,Nakama:2016gzw,Kuroyanagi:2018csn}.

PBHs are also dark matter candidate. Observations from extragalactic gamma ray background (EG$\gamma$) \cite{Carr:2009jm},
femtolensing of gamma-ray bursts \cite{Gould:1992apjl,Nemiroff:2001bp}, millilensing of compact radio sources \cite{Wilkinson:2001vv},
microlensing of quasars \cite{Dalcanton:1994apj}, the Milky way and Magellanic Cloud
stars \cite{Allsman:2000kg,Tisserand:2006zx,Griest:2013esa}
constrained the abundance of PBH dark matter \cite{Jacobs:2014yca,Sato-Polito:2019hws}.
For a recent summary of the constraints, please see Ref. \cite{Sato-Polito:2019hws}.
These constraints can be used to probe the primordial curvature perturbations at small scales. The large curvature perturbations may behave like a narrow peak \cite{Frampton:2010sw}
which can be parameterized as a delta or Gaussian function,
or an extended dome shape \cite{Garcia-Bellido:2017mdw,Gong:2017qlj,Espinosa:2017sgp,Espinosa:2018eve}.
Near the inflection point of the potential of the inflaton,
ultra-slow-roll inflation may
enhance the curvature perturbations by several orders of magnitude at small scales,
and the power spectrum for curvature perturbations remains nearly scale invariant \cite{Yi:2017mxs,Gong:2017qlj,Gao:2019sbz}.
Furthermore, an extremely blue power-law power spectrum may be generated from
axionlike curvaton inflationary model at small scales \cite{Kawasaki:2012wr}.
Therefore, the power spectrum may be parameterized as power-law form at both large and small scales.
In this paper, we propose a piecewise power-law parametrization for the power spectrum of primordial curvature perturbations,
and use the constraints on the abundance of PBH dark matter to obtain an upper limit on the power spectrum at small scales.
With the power spectrum, we calculate the secondary GWs induced by the large density perturbations at small scales.
The induced GWs can be tested by space based GW observatory like Laser Interferometer Space Antenna (LISA) \cite{Danzmann:1997hm,Audley:2017drz}, TianQin \cite{Luo:2015ght}
and TaiJi \cite{Hu:2017mde}, and the Pulsar Timing Array (PTA) \cite{Kramer:2013kea,Hobbs:2009yy,McLaughlin:2013ira,Hobbs:2013aka}
including the Square Kilometer Array (SKA) \cite{Moore:2014lga} in the future.
For simple test, we compare the strength of induced GWs with the sensitivity curves of the space based detectors \cite{Cornish:2018dyw,Liang:2019pry,Zhang:2019oet}.
On the other hand, the observations of induced GWs can also be used to constrain the power spectrum.

This paper is organized as follows. In section 2, we review the computation of the energy density of induced GWs and derive the formula for the induced GWs produced after the horizon reentry.
We propose a piecewise power-law parametrization for the power spectrum of primordial curvature perturbation in section 3,
and use current observations on PBH dark matter to obtain an upper bound on the power spectrum.
Then we use the formula derived in section 2 and the upper bound to calculate the induced GWs and discuss the possible detection of the induced GWs by future GW observations.
The conclusions are drawn in section 4.

\section{The induced GWs}\label{sec2}

Working in the Newtonian gauge, we write the perturbed metric as
\begin{equation}
    d s^2=a^2(\eta)\sqbra{-(1+2\Phi)d\eta^2+\left\{(1-2\Phi)\delta_{ij}+\frac12h_{ij}\right\}d x^i d x^j},
\end{equation}
where the scalar perturbation $\Phi$ is the Bardeen potential.
The Fourier component of the tensor perturbation $h_{ij}$ is
\begin{equation}
\label{hijkeq1}
h_{ij}(\bm{x},\eta)=\frac{1}{(2\pi)^{3/2}}\int d^3k e^{i\bm{k}\cdot\bm{x}}[h_{\bm{k}}(\eta)e_{ij}(\bm{k})+\tilde{h}_{\bm{k}}(\eta)\tilde{e}_{ij}(\bm{k})],
\end{equation}
where the plus and cross polarization tensors $e_{ij}(\bm{k})$ and $\tilde{e}_{ij}(\bm{k})$ are
\begin{equation}
\label{hijkeq2}
\begin{split}
e_{ij}(\bm{k})&=\frac{1}{\sqrt{2}}[e_i(\bm{k})e_j(\bm{k})-\tilde{e}_i(\bm{k})\tilde{e}_j(\bm{k})],\\
\tilde{e}_{ij}(\bm{k})&=\frac{1}{\sqrt{2}}[e_i(\bm{k})\tilde{e}_j(\bm{k})+\tilde{e}_i(\bm{k})e_j(\bm{k})],
\end{split}
\end{equation}
the orthonormal basis vectors $\bm{e}$ and $\tilde{\bm{e}}$ are orthogonal to $\bm{k}$, $\bm{e}\cdot \tilde{\bm{e}}=\bm{e}\cdot \bm{k}=\tilde{\bm{e}}\cdot \bm{k}=0$.
The Fourier component of the Bardeen potential $\Phi_{\bm{k}}$ is related with the primordial value $\phi_{\bm{k}}$ by the transfer function $\Psi(k\eta)$
\begin{equation}
\label{phikeq1}
    \Phi_{\bm{k}}(\eta)=\phi_{\bm{k}}\Psi(k\eta).
\end{equation}
The primordial value $\phi_{\bm{k}}$ is determined by the primordial curvature perturbation $\mathcal{P}_\zeta(k)$ as
\begin{equation}
\label{phikeq4}
\langle\phi_{\bm{k}}\phi_{\tilde{\bm{k}}}\rangle=\delta^{(3)}(\bm{k}+\tilde{\bm{k}})\frac{2\pi^2}{k^3}\left(\frac{3+3w}{5+3w}\right)^2\mathcal{P}_\zeta(k),
\end{equation}
where $w$ is determined by the time when the perturbations reenter the horizon. In this paper,
we are interested in those scales that reenter the horizon during radiation domination, so we take $w=1/3$.
During radiation domination, the transfer function is
\begin{equation}
\label{transfer}
    \Psi(x)=\frac{9}{x^2}\sbra{\frac{\sin(x/\sqrt{3})}{x/\sqrt{3}}-\cos(x/\sqrt{3})}.
\end{equation}

To the first order, the scalar perturbation decouples from tensor perturbations $h_{ij}$,
and the cosmological equation for $h_{ij}$ is homogeneous.
But to the second order, they are coupled.
The equation for induced GWs with either polarization in Fourier space with $\Phi_{\bm{k}}$ being the source is given by
\begin{equation}
\label{tensoreq}
    h_{\bm{k}}''+2\mathcal{H}h_{\bm{k}}'+k^2h_{\bm{k}}=4S_{\bm{k}},
\end{equation}
where $\mathcal{H}=a'/a$ is the conformal Hubble parameter and the prime denotes the derivative with respect to conformal time.
The source $S_{\bm{k}}$ is given by
\begin{equation}
\label{source}
    S_{\bm{k}}=\int\frac{d^3\tilde{k}}{(2\pi)^{3/2}}e_{ij}(\bm{k})\tilde{k}^i\tilde{k}^j\sbra{2\Phi_{\bm{\tilde{k}}}\Phi_{\bm{k}-\bm{\tilde{k}}}+\frac{4}{3(1+w)\mathcal{H}^2}\sbra{\Phi_{\tilde{\bm{k}}}'+\mathcal{H}\Phi_{\tilde{\bm{k}}}}\sbra{\Phi_{\bm{k}-\bm{\tilde{k}}}'+\mathcal{H}\Phi_{\bm{k}-\bm{\tilde{k}}}}}.
\end{equation}
The power spectrum of the induced GWs is defined as
\begin{equation}
\label{hijkeq4}
  \langle h_{\bm{k}}(\eta)h_{\tilde{\bm{k}}}(\eta)\rangle =\frac{2\pi^2}{k^3}\delta^{(3)}(\bm{k}+\tilde{\bm{k}})\mathcal{P}_{h}(k,\eta),
\end{equation}
and the fractional energy density is
\begin{equation}
\label{density}
    \Omega_{\mathrm{GW}}(k,\eta)=\frac{1}{24}\sbra{\frac{k}{aH}}^2\overline{\mathcal{P}_h(k,\eta)},
\end{equation}
where the Hubble parameter $H=\mathcal{H}/a$. Before presenting the detailed derivation of the induced GWs, we discuss its qualitative behavior first.
Following \cite{Baumann:2007zm}, we assume that the induced GWs are produced instantaneously when the relevant scales reenter the horizon. At the horizon reentry,
$h_{\bm{k}}\sim S_{\bm{k}}/k^2$ and it gets contributions from all scalar modes $\Phi_{\tilde{\bm{k}}}$. However, combining Eqs. \eqref{source} and \eqref{hijkeq4},
it is easy to see that $k^3\tilde{k}^3/|\bm{k}-\tilde{\bm{k}}|^3$ appears in
the integrand in $\mathcal{P}_h$, so the main contributions to $\mathcal{P}_h$ are from $\tilde{\bm{k}}$ that are close to $\bm{k}$. Since the source $S_{\bm{k}}$ decays as $a^{-\gamma}$ with $3\leq\gamma\leq 4$ \cite{Baumann:2007zm}, soon after the horizon reentry GWs propagate freely and $h_{\bm{k}}\propto a^{-1}$,
so $\Omega_{\text{GW}}(k,\eta)$ is a constant well within the horizon.

In terms of Green's function $G_{\bm{k}}(\eta,\tilde{\eta})$ satisfying the equation
\begin{equation}
\label{greeneq}
G_{\bm{k}}''(\eta,\tilde{\eta})+\left(k^2-\frac{a''(\eta)}{a(\eta)}\right)G_{\bm{k}}(\eta,\tilde{\eta})=\delta(\eta-\tilde{\eta}),
\end{equation}
the solution to Eq. \eqref{tensoreq} is
\begin{equation}
\label{hijkeq3}
h_{\bm{k}}(\eta)=\frac{4}{a(\eta)}\int_{\eta_k}^\eta d\tilde{\eta} G_{\bm{k}}(\eta,\tilde{\eta})a(\tilde{\eta})S_{\bm{k}}(\tilde{\eta}).
\end{equation}
Note that we assume that the induced GWs are produced after the horizon reentry, so we take $k\eta_k=1$.
During radiation domination, the Green's function is
\begin{equation}
\label{hijkeq8}
G_{\bm{k}}(\eta,\tilde{\eta})=\frac{1}{k}\sin [k(\eta-\tilde{\eta})].
\end{equation}
Combining Eqs. \eqref{phikeq1}, \eqref{transfer}, \eqref{source}, \eqref{hijkeq4} and \eqref{hijkeq3}, after a straightforward and tedious
calculation, we obtain the power spectrum of the induced GWs \cite{Ananda:2006af,Baumann:2007zm,Inomata:2016rbd,Kohri:2018awv}
\begin{equation}
\label{ph}
\mathcal{P}_{h}(k, \eta)=4 \int_{0}^{\infty} d v \int_{|1-v|}^{1+v} d u\left[\frac{4 v^{2}-\left(1-u^{2}+v^{2}\right)^{2}}{4 u v}\right]^{2} I_{\mathrm{RD}}^{2}(u, v, x) \mathcal{P}_{\zeta}(k v) \mathcal{P}_{\zeta}(k u),
\end{equation}
where $u= |\bm{k}-\tilde{\bm{k}}|/k$, $v= \tilde{k}/k$, $x= k\eta$,
the power spectrum $\mathcal{P}_\zeta(k)$ for the primordial curvature perturbation is evaluated at horizon exit during inflation.
Combining Eqs. \eqref{density}, \eqref{ph} and \eqref{averaged}, we get induced GWs in radiation dominated era,
\begin{equation}
\label{sgwres1}
\Omega_{\mathrm{GW}}(k,\eta)=\frac{1}{6}\left(\frac{k}{aH}\right)^2 \int_{0}^{\infty} d v \int_{|1-v|}^{1+v} d u\left[\frac{4 v^{2}-\left(1-u^{2}+v^{2}\right)^{2}}{4 u v}\right]^{2} \overline{I_{\mathrm{RD}}^{2}(u, v, x)} \mathcal{P}_{\zeta}(k v) \mathcal{P}_{\zeta}(k u).
\end{equation}
For the convenience of taking the time average,
we split the source term $I_{\text{RD}}$ in the radiation era into the combinations of two oscillations \cite{Cai:2019jah},
\begin{equation}
\label{irdeq1}
I_{\text{RD}}(u, v, x)=\frac{1}{9x}\sbra{I_s\sin{x}+I_c\cos{x}},
\end{equation}
where
$I_c$ and $I_s$ are given by \footnote{We learned from Davide Racco that similar results were obtained in \cite{Espinosa:2018eve}.}
\begin{equation}
\label{iceq11}
    I_{c}(u, v, x)=-4\int_1^x y \sin(y) f(y) d y=T_{c}(u, v, x)-T_{c}(u, v, 1),
\end{equation}
\begin{equation}
\label{iseq11}
    I_{s}(u, v, x)=4 \int_1^x y \cos(y) f(y) d y=T_{s}(u, v, x)-T_{s}(u, v, 1),
\end{equation}
\begin{equation}
\label{tceq1}
T_c(u, v, x)=-4\int_0^x y \sin(y) f(u,v,y) d y,
\end{equation}
\begin{equation}
\label{tseq1}
T_s(u, v, x)=4 \int_0^x y \cos(y) f(u,v,y) d y,
\end{equation}
and
\begin{equation}
\label{irdeq1a}
f(u,v,x)=2 \Psi(v x) \Psi(u x)+\left[\Psi(v x)+vx\Psi'(v x)\right]\left[\Psi(u x)+ux\Psi'(u x)\right].
\end{equation}
Note that we take the point of view that induced GWs are produced after the relevant modes reenter the horizon, the lower limit of the integrals \eqref{iceq11} and \eqref{iseq11} should be 1 \cite{Baumann:2007zm,Espinosa:2018eve},
so we need to subtract the terms $T_c(u,v,1)$ and $T_s(u,v,1)$ in Eqs. \eqref{iceq11} and \eqref{iseq11}.
In \cite{Inomata:2016rbd,Kohri:2018awv}, the lower limit of the integrals \eqref{iceq11} and \eqref{iseq11} was chosen to be zero, i.e.,
it was assumed that the production of induced GWs begins long before the horizon reentry. We expect the choice of the lower limit of the integral will affect small $k$ modes more because they stay outside the horizon longer.
If we take $I_c(u,v,x)=T_c(u,v,x)$ and $I_s(u,v,x)=T_s(u,v,x)$,
then we recover the result for $I_{\text{RD}}(u, v, x)$ in \cite{Kohri:2018awv}.

Substituting the transfer function \eqref{transfer} into Eqs. \eqref{tceq1} and \eqref{tseq1}, we get
\begin{equation}
\label{tceq2}
\begin{split}
    T_c=&\frac{-27}{8u^3v^3x^4}\left[\vphantom{\frac12}-48uvx^2(x\cos x+3\sin x)\cos\frac{ux}{\n3}\cos\frac{vx}{\n3}\right.\\
    &+48\n3 x^2\cos{x}\sbra{v\cos{\vx}\sin{\ux}+u\cos{\ux}\sin{\vx}}\\
    &+8\n3 x\sin{x}\left(\vphantom{\frac12}\right.[18-x^2(u^2+3-v^2)]v\cos{\vx}\sin{\ux}\\
    &+[18-x^2(v^2+3-u^2)]u\cos{\ux}\sin{\vx}\left.\vphantom{\frac12}\right)\\
    &+24x[-6+x^2(3-u^2-v^2)]\cos{x}\sin{\ux}\sin{\vx}\\
    &\left.+24[-18+x^2(3+u^2+v^2)]\sin{x}\sin{\ux}\sin{\vx}\right]\\
    &-\frac{27(u^2+v^2-3)^2}{4u^3v^3}\left(\Si\left[\left(1-\frac{u-v}{\sqrt{3}}\right)x\right]
    +\Si\left[\left(1+\frac{u-v}{\sqrt{3}}\right)x\right]\right.\\
    &\left.\qquad\qquad -\Si\left[\left(1-\frac{u+v}{\sqrt{3}}\right)x\right]-
    \Si\left[\left(1+\frac{u+v}{\sqrt{3}}\right)x\right]\right),
\end{split}
\end{equation}
and
\begin{equation}
\label{tseq2}
\begin{split}
    T_s=&\frac{27}{8u^3v^3x^4}\left[\vphantom{\frac12}
    48uvx^2(x\sin x-3\cos x)\cos\frac{ux}{\n3}\cos\frac{vx}{\n3}\right.\\
    &-48\n3 x^2\sin{x}\sbra{v\cos{\vx}\sin{\ux}+u\cos{\ux}\sin{\vx}}\\
    &+8\n3 x\cos{x}\left(\vphantom{\frac12}\right.[18-x^2(u^2+3-v^2)]v\cos{\vx}\sin{\ux}\\
    &+[18-x^2(v^2+3-u^2)]u\cos{\ux}\sin{\vx}\left.\vphantom{\frac12}\right)\\
    &+24x[6-x^2(3-u^2-v^2)]\sin{x}\sin{\ux}\sin{\vx}\\
    &\left.+24[-18+x^2(3+u^2+v^2)]\cos{x}\sin{\ux}\sin{\vx}\right]-\frac{27(u^2+v^2-3)}{u^2v^2}\\
    &+\frac{27(u^2+v^2-3)^2}{4u^3 v^3}\left(\Ci\left[\left(1-\frac{u-v}{\sqrt{3}}\right)x\right]
    +\Ci\left[\left(1+\frac{u-v}{\sqrt{3}}\right)x\right]\right.\\
    &\left.\qquad\qquad -\Ci\left[\left|1-\frac{u+v}{\sqrt{3}}\right|x\right]-
    \Ci\left[\left(1+\frac{u+v}{\sqrt{3}}\right)x\right]+\ln\left|\frac{3-(u+v)^2}{3-(u-v)^2}\right|\right).
\end{split}
\end{equation}
The sine-integral function Si$(x)$ and cosine-integral function Ci$(x)$ are defined as
\begin{align}
    \text{Si}(x)=\int_0^x d y\frac{\sin y}{y},\quad \text{Ci}(x)=-\int_x^\infty d y \frac{\cos y}{y}.
\end{align}
They have the asymptotic behavior $\text{Si}(x)\rightarrow x$ and $\text{Ci}(x)\rightarrow \ln(x)+\gamma_E$ as $x\rightarrow 0$,
here $\gamma_E$ is the Euler number.

At late times, $\eta\gg \eta_k$ and $x\rightarrow \infty$,
\begin{equation}
\label{hijkeq16}
\begin{split}
    I_{\text{RD}}(u,v,x\rightarrow\infty)=&-\frac{3\pi(u^2+v^2-3)^2\Theta(u+v-\n3)}{4u^3v^3x}\cos{x}\\
    &-\frac{1}{9x}\sbra{T_c(u,v,1)\cos{x}+\tilde{T}_s(u,v,1)\sin{x}},
\end{split}
\end{equation}
where
\begin{equation}
\label{hijkeq16a}
\tilde{T}_s(u,v,1)=T_s(u,v,1)+\frac{27(u^2+v^2-3)}{u^2v^2}-\frac{27(u^2+v^2-3)^2}{4u^3 v^3}\ln\left|\frac{3-(u+v)^2}{3-(u-v)^2}\right|.
\end{equation}
So the time average is
\begin{equation}
\label{averaged}
\begin{split}
    \overline{I^2_{\text{RD}}(u,v,x\rightarrow\infty)}=&\frac{1}{2x^2}\left[\left(\frac{3\pi(u^2+v^2-3)^2\Theta(u+v-\n3)}{4u^3v^3}+\frac{T_c(u,v,1)}{9}\right)^2\right.\\
    &\qquad \left.+\left(\frac{\tilde{T}_s(u,v,1)}{9}\right)^2\right].
\end{split}
\end{equation}
Substituting \eqref{averaged} into \eqref{ph}, we find that $\overline{\mathcal{P}_h(k,\eta)}\sim 1/\eta^2$ for
the modes well inside the horizon in
the radiation dominated era.
During radiation domination,
$\mathcal{H}=aH\sim 1/\eta$, so $\Omega_{\text{GW}}$ is time independent late in the radiation dominated era as discussed above.
Since GWs behave like radiation, the current energy densities of GWs
are related to their values well after the horizon reentry in the radiation
dominated era
\begin{equation}
\label{propagation}
\Omega_{\text{GW}}\left(k, \eta_{0}\right)=\Omega_{\text{GW}}(k, \eta) \frac{\Omega_{r0}}{\Omega_{r}(\eta)},
\end{equation}
where $\Omega_r$ is the fractional energy density of radiation,
$\eta\gg \eta_k$ is chosen to be earlier than the matter radiation equality
and late enough so that $\Omega_{\text{GW}}(k, \eta)$ is a constant, and the subscript $0$ denotes for quantities evaluated at today.

Once we are given the power spectrum $\mathcal{P}_\zeta(k)$ for the primordial curvature perturbation,
we combine Eqs. \eqref{sgwres1} and \eqref{averaged} to calculate induced GWs in radiation dominated era,
and obtain $\Omega_{\text{GW}}(k,\eta_0)$ from Eq. \eqref{propagation}. In the following, we use several examples to calculate $\Omega_{\text{GW}}$.

\subsection{The scale invariant power spectrum}
For the scale invariant power spectrum, $\mathcal{P}_\zeta(k)=A_\zeta$,
the numerical integration gives
\begin{equation}
\Omega(k,\eta)\approx 0.7859 A_\zeta^2.
\end{equation}
Comparing with the result $\Omega(k,\eta)\approx 0.8222 A_\zeta^2$ obtained in \cite{Kohri:2018awv}
by assuming that the production of induced GWs starts well before the horizon reentry, this value is about 4.6\% smaller, so the contribution by the induced GWs produced before the horizon reentry is small. In this case, all the modes contribute
equally in the integration \eqref{sgwres1} and the contribution to the total integral by $T_c(u,v,1)$ and $T_s(u,v,1)$ in Eq. \eqref{irdeq1} is small.

\subsection{The power-law power spectrum}
\begin{figure}[htp]
    \centering
    \includegraphics[width=120mm]{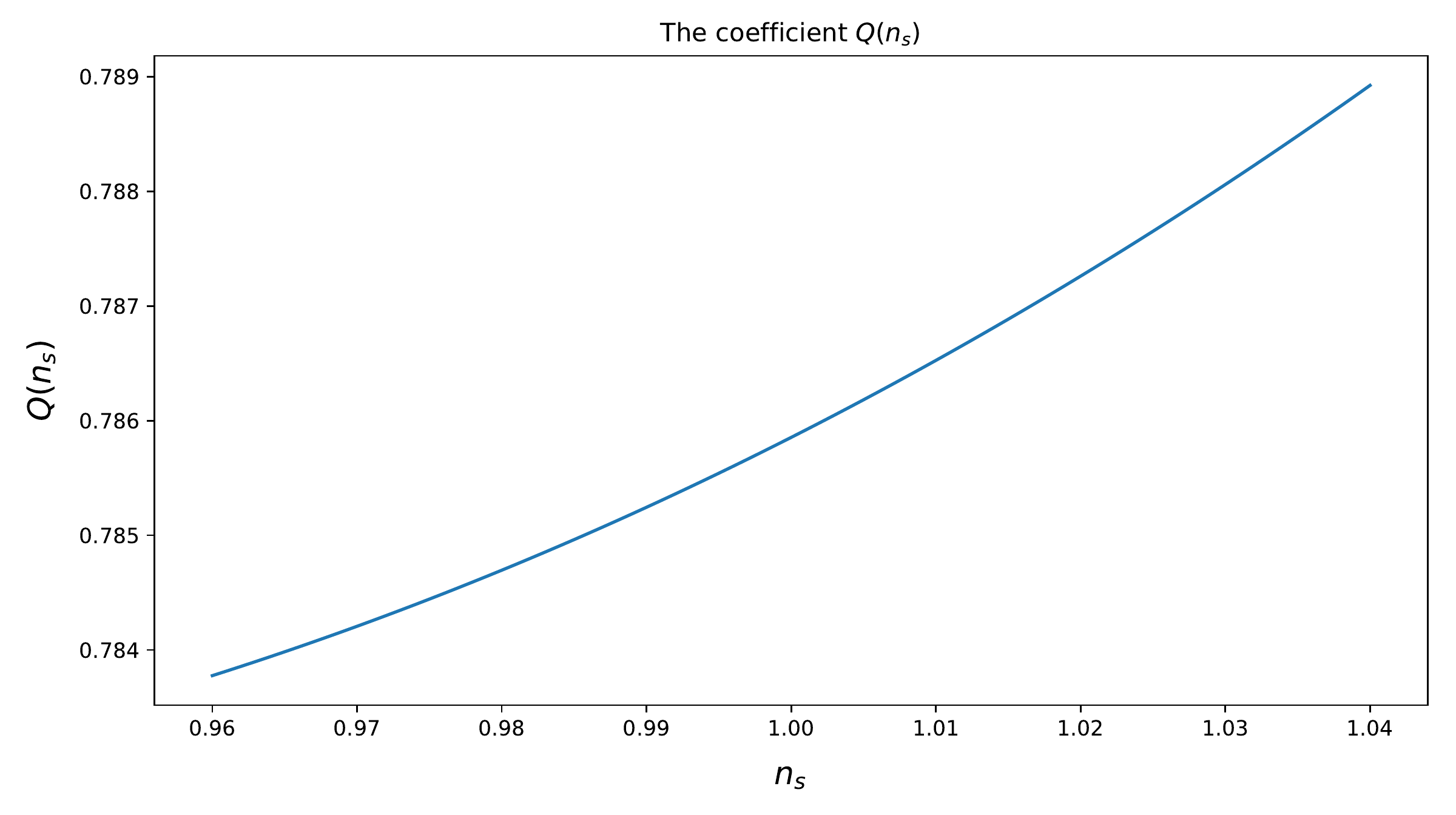}
    \caption{The value of $Q(n_s)$ as a function of $n_s$.}
    \label{fig:Qns}
\end{figure}

For a nearly scale invariant power spectrum with the power-law form,
\begin{equation}
    \mathcal{P}_\zeta(k)=A_\zeta\sbra{\frac{k}{k_p}}^{n_s-1},
\end{equation}
we get
\begin{equation}\label{powerlawGW}
    \Omega_{\mathrm{GW}}(k,\eta)=Q(n_s)A_\zeta^2\sbra{\frac{k}{k_p}}^{2(n_s-1)},
\end{equation}
where the factor $Q(n_s)$ needs to be calculated numerically. We show the numerical results for $Q(n_s)$ in Fig. \ref{fig:Qns}. Again, the results are about 5\% smaller than those in \cite{Kohri:2018awv}. Similar to the scale invariant case,
all the modes contribute in the integration \eqref{sgwres1}
and the contribution to the total integral by $T_c(u,v,1)$ and $T_s(u,v,1)$ in Eq. \eqref{irdeq1} is small.
In \cite{Baumann:2007zm},
it was estimated that $Q(n_s)\approx 10$, so that estimate is an order of magnitude larger than the more accurate result $Q(n_s)\approx 0.8$.

\subsection{The monochromatic power spectrum}
\begin{figure}[htp]
  \centering
  \includegraphics[width=120mm]{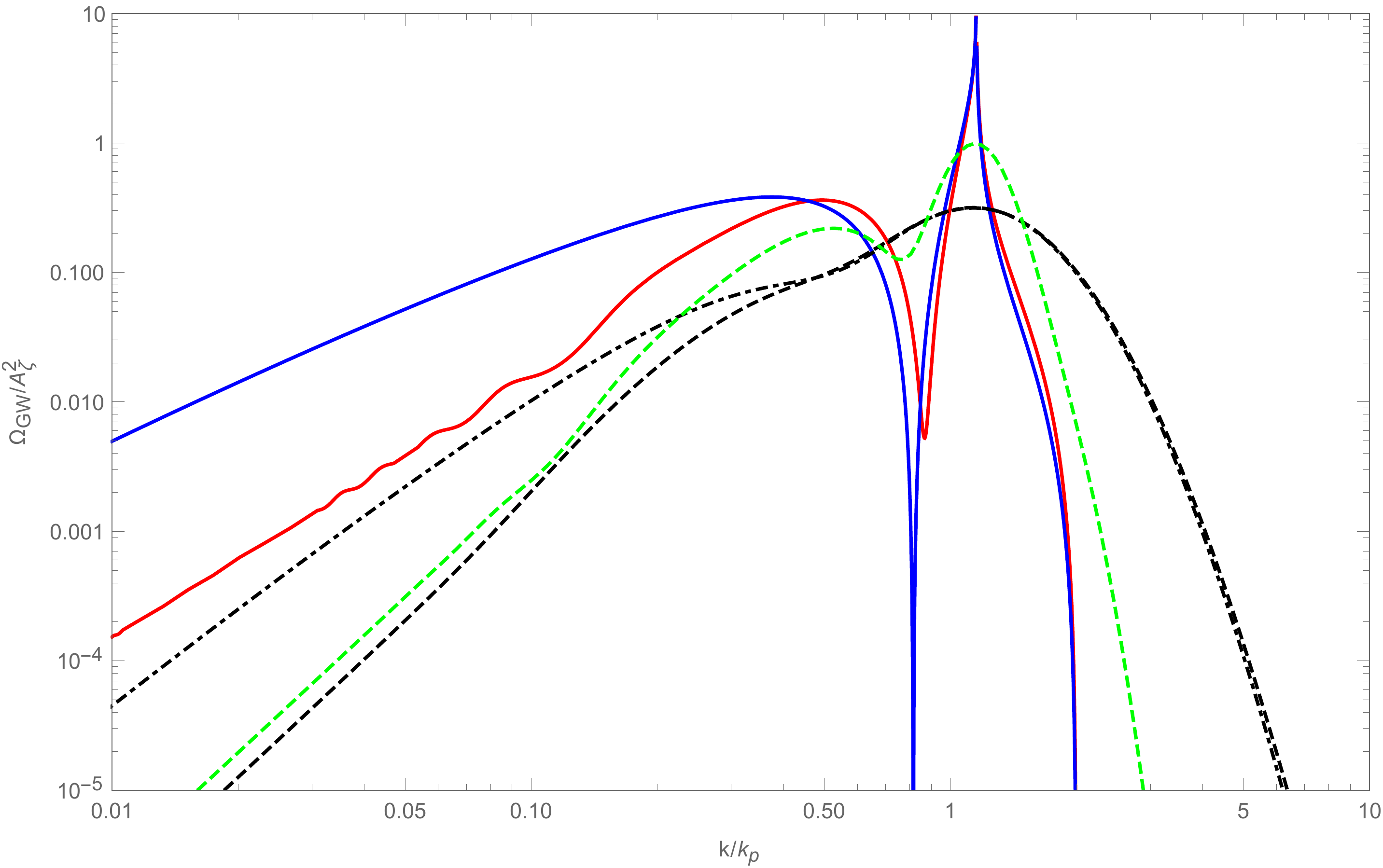}
  \caption{The induced GWs ($\Omega_GW/A_\zeta^2$)
  from monochromatic and Gaussian power spectra of curvature perturbations.
  The red solid line denotes the induced GWs from the monochromatic power spectrum,
  the green and black dashed lines denote the induced GWs from the Gaussian power spectrum with $\sigma=0.2$ and $\sigma=0.5$, respectively.
  For comparison, we also show the induced GWs from
  the monochromatic power spectrum and Gaussian power spectrum with $\sigma=0.5$ by using the formulae derived in \cite{Kohri:2018awv} with the blue solid line
  and the black dot dashed line, respectively.}
  \label{mono_comp}
\end{figure}

Now we consider induced GWs generated by the monochromatic
curvature perturbations with the delta-function-type power spectrum
which can be taken as the idealized limit of a peak in the power spectrum \cite{Ananda:2006af,Inomata:2016rbd,Kohri:2018awv}
\begin{equation}\label{monodelta}
    \mathcal{P}_\zeta(k)=A_\zeta\delta\sbra{\ln{\frac{k}{k_p}}},
\end{equation}
where $A_\zeta$ is the amplitude and $k_p$ is the wave number
at which the delta-function peak occurs.
Plugging Eq. \eqref{monodelta} into Eq. \eqref{sgwres1},
we get the corresponding induced GWs
\begin{equation}
\label{monpweq}
\begin{split}
    \Omega_{\text{GW}}&=A_\zeta^2\times \frac{\bar{k}^2}{192}\left(\frac{4}{\bar{k}^2}-1\right)^2\Theta(2-\bar{k})\left[\left(\frac{\tilde{T}_s(\bar{k}^{-1},\bar{k}^{-1},1)}{9}\right)^2\right.\\
    &\left.\qquad +\left(\frac{3\bar{k}^6\pi}{4}\sbra{\frac{2}{\bar{k}^2}-3}^2\Theta(2-\n3\bar{k})+\frac{T_c(\bar{k}^{-1},\bar{k}^{-1},1)}{9}\right)^2\right],\\
    &=A_\zeta^2\times\frac{3\bar{k}^6}{1024}\left(1-\frac{4}{\bar{k}^2} \right)^2\Theta(2-\bar{k})\left\{\vphantom{\frac12}\left[A(\bar{k})+(2-3\bar{k}^2)^2\pi\Theta(2-\sqrt{3}\bar{k})\right]^2\right.\\
  &\left.\qquad+\left[B(\bar{k})+(2-3\bar{k}^2)^2\left(2\Ci(1)-\Ci\left(1+\frac{2}{\sqrt{3}\bar{k}}\right)-\Ci\left(\left|1-\frac{2}{\sqrt{3}\bar{k}}\right|\right)\right) \right]^2\right\},
\end{split}
\end{equation}
where $\bar{k}\equiv k/k_p$,
\begin{equation}
\label{Ak}
\begin{split}
  A(k)=&24 k^2[3\sin (1)+\cos (1)]-12 \sqrt{3} k^3 [5\sin (1)+2\cos (1)] \sin \left(\frac{2}{\sqrt{3} k}\right)\\
  &+12 k^2 [3 k^2\cos (1) + \left(15 k^2-8\right)\sin (1)]
  \sin^2\left(\frac{1}{\sqrt{3} k}\right) \\
  &+\left(2-3 k^2\right)^2\left(-2 \text{Si}(1)+ \text{Si}\left(1+\frac{2}{\sqrt{3} k}\right)+ \text{Si}\left(1-\frac{2}{\sqrt{3} k}\right)\right),
\end{split}
\end{equation}
and
\begin{equation}
\label{Bk}
\begin{split}
  B(k)=&24 k^2[\sin (1)-3\cos (1)]+12\sqrt{3} k^3[5\cos (1)-2 \sin (1)] \sin \left(\frac{2}{\sqrt{3} k}\right)\\
  &+12k^2 [3k^2 \sin (1) + \left(8-15 k^2\right)\cos (1) ]
  \sin^2\left(\frac{1}{\sqrt{3} k}\right).
\end{split}
\end{equation}
The result for Eq. \eqref{monpweq} is shown as the red solid line in Fig. \ref{mono_comp}.
In the integral \eqref{sgwres1}, only the mode $u=v=\bar{k}^{-1}$
contributes to the integration, so $\Omega_{GW}$ is determined
by Eq. \eqref{averaged} with $u=v=\bar{k}^{-1}$. Around $\bar{k}=2/\sqrt{3}$,
the last $\Ci$ term in \eqref{monpweq} is logarithmic divergent,
so there is a sharp peak
in $\Omega_{\text{GW}}$ at $\bar{k}=2/\sqrt{3}$ as shown in Fig. \ref{mono_comp}.
Before the sharp peak, some terms in \eqref{monpweq} may cancel each other,
so there is a dip in $\Omega_{\text{GW}}$ before the sharp peak.
The sharp peak at $k=2k_p/\sqrt{3}$ is due to the resonant amplification \cite{Ananda:2006af,Kohri:2018awv}.
The factor 2 is from the second order effect (the source is $\Phi^2$)
and the factor $1/\sqrt{3}$ is due to the sound speed of radiation background.
For comparison, the blue solid line in Fig. \ref{mono_comp} shows the induced GWs from the monochromatic power spectrum obtained in \cite{Kohri:2018awv}.
The difference comes from the generation of GWs before the horizon reentry.
In \cite{Kohri:2018awv}, they assume that the production of induced GWs starts
when the Universe begins radiation domination. In deriving Eq. \eqref{monpweq},
we assume that induced GWs are produced after the relevant scales reenter the horizon.
For the monochromatic power spectrum, only single mode $u=v=\bar{k}^{-1}$
contributes to the integration, and large scales reenter the horizon
at later time, so the contribution by the production of induced GWs well outside
the horizon will be larger. As shown in Fig. \ref{mono_comp}, the
difference can be several orders of magnitude at large
scales ($k/k_p\lesssim 0.1$), but the
difference at small scales ($k/k_p\gtrsim 1$) is small.

\subsection{The Gaussian power spectrum}

More generally, the peak in the power spectrum may be
parameterized as Gaussian form \cite{Cai:2018dig,Inomata:2018epa}
\begin{equation}
\label{GaussD}
  \mathcal{P}_{\zeta}(k)=\frac{A_\zeta}{\sigma\sqrt{2\pi}}\exp\left(-\frac{\ln^2(k/k_p)}{2\sigma^2} \right).
\end{equation}
In the limit $\sigma\rightarrow 0$, the sharp peak with delta function \eqref{monodelta} is recovered.
The smaller width $\sigma$ corresponds to a sharper peak on power spectrum.
Combining Eqs. \eqref{sgwres1}, \eqref{averaged} and \eqref{GaussD},
we calculate induced GWs produced by Gaussian power spectrum
with $\sigma=0.2$ and $\sigma=0.5$,
and the results are shown as the dashed green and black lines in Fig. \ref{mono_comp}.
Since more modes contribute to the integration \eqref{sgwres1},
the resonance peak is broadened and the amplitude is reduced as $\sigma$ becomes larger. The dip before the peak disappears when $\sigma$ is large enough.
We also show the induced GWs by considering the production of GWs
before the horizon reentry with the black dot dashed line in Fig. \ref{mono_comp}.
The difference due to the production of GWs before the horizon reentry is large at large scales.

\section{PBH and the observational constraints}\label{sec3}
PBHs form in the region with its density contrast at horizon reentry exceeding the threshold $\delta_c$.
Suppose the density perturbations are Gaussian, the probability distribution of the smoothed density contrast $\delta(R)$ over a sphere with comoving radius $R$ is  \cite{Liddle:2000cg}
\begin{equation}
\label{pbheq1}
    P(\delta(R))=\frac{1}{\sqrt{2\pi\sigma^2(R)}}\exp\sbra{-\frac{\delta^2(R)}{2\sigma^2(R)}},
\end{equation}
where the smoothing scale $R$ is the horizon size, $R=\mathcal{H}^{-1}$ and the mass variance $\sigma(R)$ associated with the PBH mass $M_\text{PBH}$ is
\begin{equation}
\label{pbheq2}
    \sigma^2(R)=\int_0^\infty {W}^2(kR)\frac{\mathcal{P}_\delta(k)}{k} d k,
\end{equation}
$\mathcal{P}_\delta$ is the power spectrum of the matter perturbation and the window function is ${W}(kR)=\exp\sbra{-k^2R^2/2}$.
During radiation domination, the matter perturbation relates to the primordial curvature perturbation as
\begin{equation}
\label{pbheq3}
    \mathcal{P}_\delta(k)=\frac{16}{81}\sbra{\frac{k}{aH}}^4\mathcal{P}_{\zeta}(k).
\end{equation}
Using Press-Schechter theory \cite{Press:1973iz}, we get the fraction of the energy density in the Universe going to PBHs \footnote{There should be a
factor $\gamma$ in \eqref{7} \cite{Alabidi:2012ex}. However, it has very little effect on the result, so we ignore this factor here.}
\begin{equation}
\label{7}
    \beta(M_{\text{PBH}})=2\int_{\delta_c}^\infty P(\delta)d \delta
    =\t{erfc}\sbra{\frac{\delta_c}{\sqrt{2}\sigma}},
\end{equation}
where $\delta_c=0.42$ \cite{Harada:2013epa}. Combining Eqs. \eqref{pbheq2} and \eqref{pbheq3}, we see that
the dominant contribution to the mass variance $\sigma^2(R)$
comes from the scale $k=1/R$, so $\sigma^2(R)\propto \mathcal{P}_\zeta(1/R)$.
Following Ref. \cite{Sato-Polito:2019hws}, at each $k$, we calculate $\sigma^2(R)$ with scale invariant $\mathcal{P}_\zeta$,
so we have
\begin{equation}
\label{pbheq4}
\beta \approx \text{erfc}\left(\frac{9\delta_c}{4\sqrt{\mathcal{P}_\zeta}}\right).
\end{equation}

Since PBH forms in the radiation dominated era, the mass of PBH is of the order of the horizon mass $M_H=4\pi \rho/(3H^3)=(2GH)^{-1}$ \cite{Nakama:2016gzw}
\begin{equation}
\label{pbhmass}
    M_{\text{PBH}}=\gamma M_H=\gamma\left.\Omega_{\text{r}0}^{1/2}M_0\sbra{\frac{g_*^0}{g_*^i}}^{1/6}\sbra{\frac{H_0}{k}}^2\right|_{k=aH},
\end{equation}
where the order one ratio $\gamma$ is chosen as $\gamma=3^{-3/2}\approx 0.2$ \cite{Carr:1975qj},
$\Omega_{\text{r}0}=9.17 \times 10^{-5}$, $M_0=(2GH_0)^{-1}\approx 4.63\times 10^{22}~M_{\odot}$, $H_0=67.27 \text{ km/s/Mpc}$ \cite{Aghanim:2018eyx},
$g_*^0\approx 3.36$ and $g_*^i$ denote the effective degrees of freedom for energy density at present and at the formation of PBH respectively.
In this paper, we don't distinguish the difference between the effective degrees of freedom for the entropy and energy density.
For the mass scale of PBHs we are interested in, we take $g_*^i\approx 10.75$.
After their formation, PBHs behave like matter, so the energy fraction of PBHs increases until the matter radiation equality.
Ignoring the mass accretion and evaporation,
the energy fraction of PBHs at their formation is
\begin{equation}
\label{3}
    \beta(M_\text{PBH})= 4\times 10^{-9}\sbra{\frac{\gamma}{0.2}}^{-1/2}\sbra{\frac{g_*^i}{10.75}}^{1/4}\sbra{\frac{M_\text{PBH}}{M_\odot}}^{1/2}f_\text{PBH},
\end{equation}
where $f_\text{PBH}=\Omega_\text{PBH}/\Omega_{\t{DM}}$ is the current energy fraction of PBHs $\Omega_\text{PBH}$ to dark matter $\Omega_{\t{DM}}$.

\begin{figure}[htp]
    \centering
    \includegraphics[width=120mm]{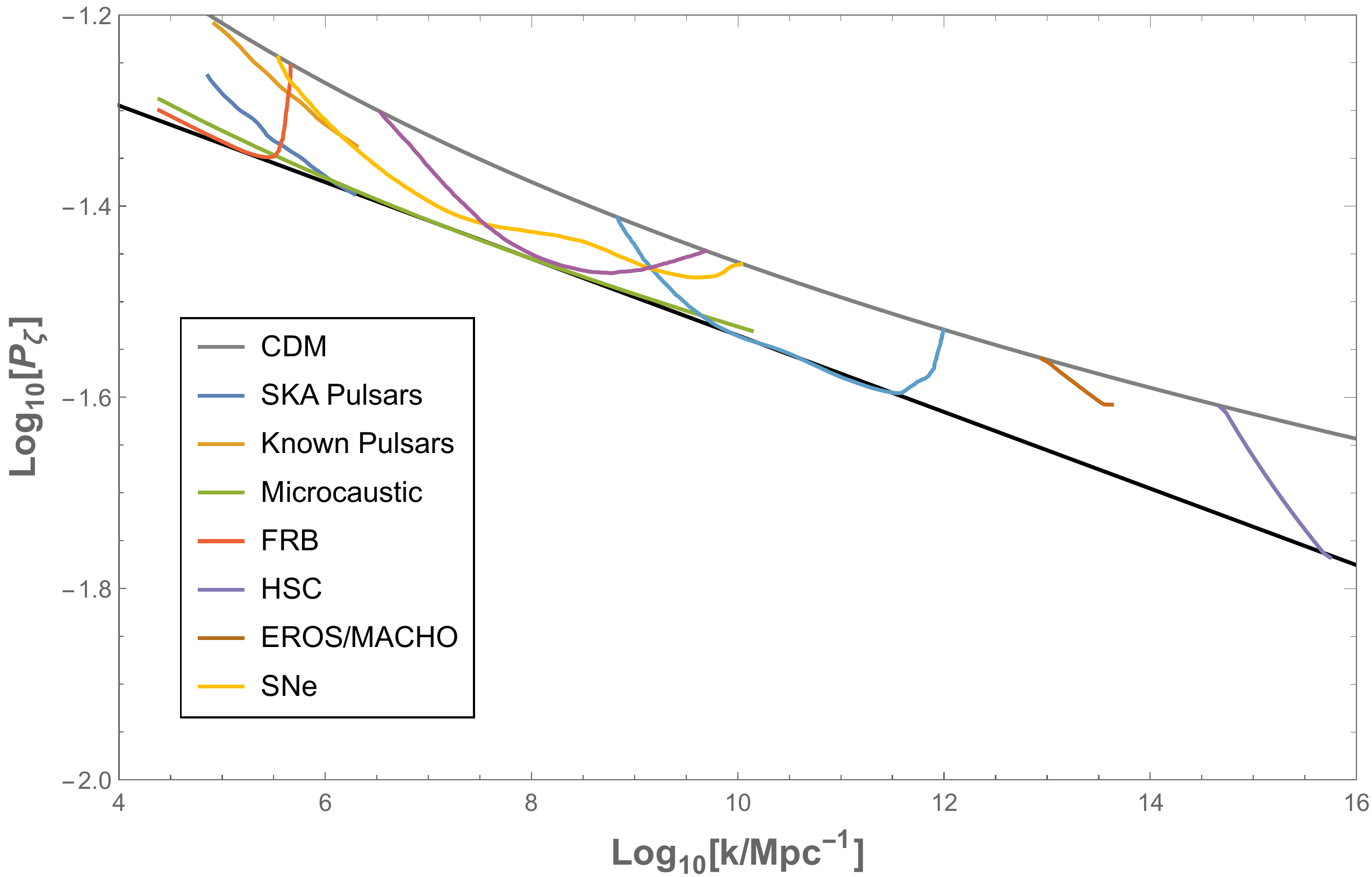}
    \caption{The observational constraints on the power spectrum of primordial curvature perturbations.
For the details of observational constraints, please refer to \cite{Sato-Polito:2019hws} and references therein.
The solid black line is the upper limit obtained by the piecewise power-law parametrization \eqref{pzeq3}.}
    \label{fig:PBHonPowerSpec}
\end{figure}

Combining Eqs. \eqref{pbheq4} and \eqref{3}, we can obtain $\mathcal{P}_{\zeta}$ for a given $f_{\t{PBH}}$ and vice versa.
This allows us to use the observational constraints on PBH abundance, namely $f_{\t{PBH}}$,
to constrain the power spectrum for primordial curvature perturbations at small scales.
Alternatively, it allows us to use $f_{\t{PBH}}$ to constrain some inflationary models.
The current observational constraints on $f_{\text{PBH}}$ and $\mathcal{P}_\zeta$ at small scales
were summarized in Ref. \cite{Sato-Polito:2019hws} and we show them in Fig. \ref{fig:PBHonPowerSpec}.
On observable scales $10^{-4}\text{ Mpc}^{-1}\lesssim k\lesssim 10^{-1}\text{ Mpc}^{-1}$, the temperature and polarization
measurements on the cosmic microwave background anisotropy constrain the nearly scale invariant power spectrum for the primordial curvature perturbation as \cite{Akrami:2018odb}
\begin{equation}
\label{pzeq1}
    \mathcal{P}_{\zeta}=A_s \left(\frac{k}{k_*}\right)^{n_s-1},
\end{equation}
where $k_*=0.05 \text{ Mpc}^{-1}$, $A_s=2.1\times 10^{-9}$ and $n_s=0.9649\pm 0.0044$.

To get large enhancement on the power spectrum from a single field inflation,
an ultra-slow-roll inflation near the inflection point may be used,
and the power spectrum is also nearly scale invariant \cite{Yi:2017mxs,Gong:2017qlj,Gao:2019sbz}.
Furthermore, an extremely blue power-law power spectrum may be generated from
axionlike curvaton model at small scales \cite{Kawasaki:2012wr}. For simplicity,
we use the piecewise power-law parametrization
for the power spectrum to fit the observational
bounds in Fig. \ref{fig:PBHonPowerSpec} and the fitting results are
\begin{equation}
\label{pzeq3}
    \mathcal{P}_{\zeta}(k)=\left\{
    \begin{aligned}
        &2.1\times 10^{-9}\sbra{\frac{k}{0.05\t{ Mpc}^{-1}}}^{0.9649-1},\quad  &k\lesssim 1\t{ Mpc}^{-1}\\
        &2.1\times 10^{-9}\sbra{\frac{k}{0.05 \text{ Mpc}^{-1}}}^{1.857},\quad  &1\t{ Mpc}^{-1}\lesssim k\lesssim 10^4\t{ Mpc}^{-1}\\
        &5.1\times 10^{-2} \sbra{\frac{k}{10^4\t{ Mpc}^{-1}}}^{0.960-1},\quad  & k\gtrsim 10^4\t{ Mpc}^{-1}
    \end{aligned}
    \right.
\end{equation}
We show this piecewise power-law parametrization of the power spectrum in Fig. \ref{fig:PBHonPowerSpec} by the solid black line.
The power spectrum well fits the upper bound from PBHs and the CMB constraints.

\begin{figure}[htp]
    \centering
    \includegraphics[width=120mm]{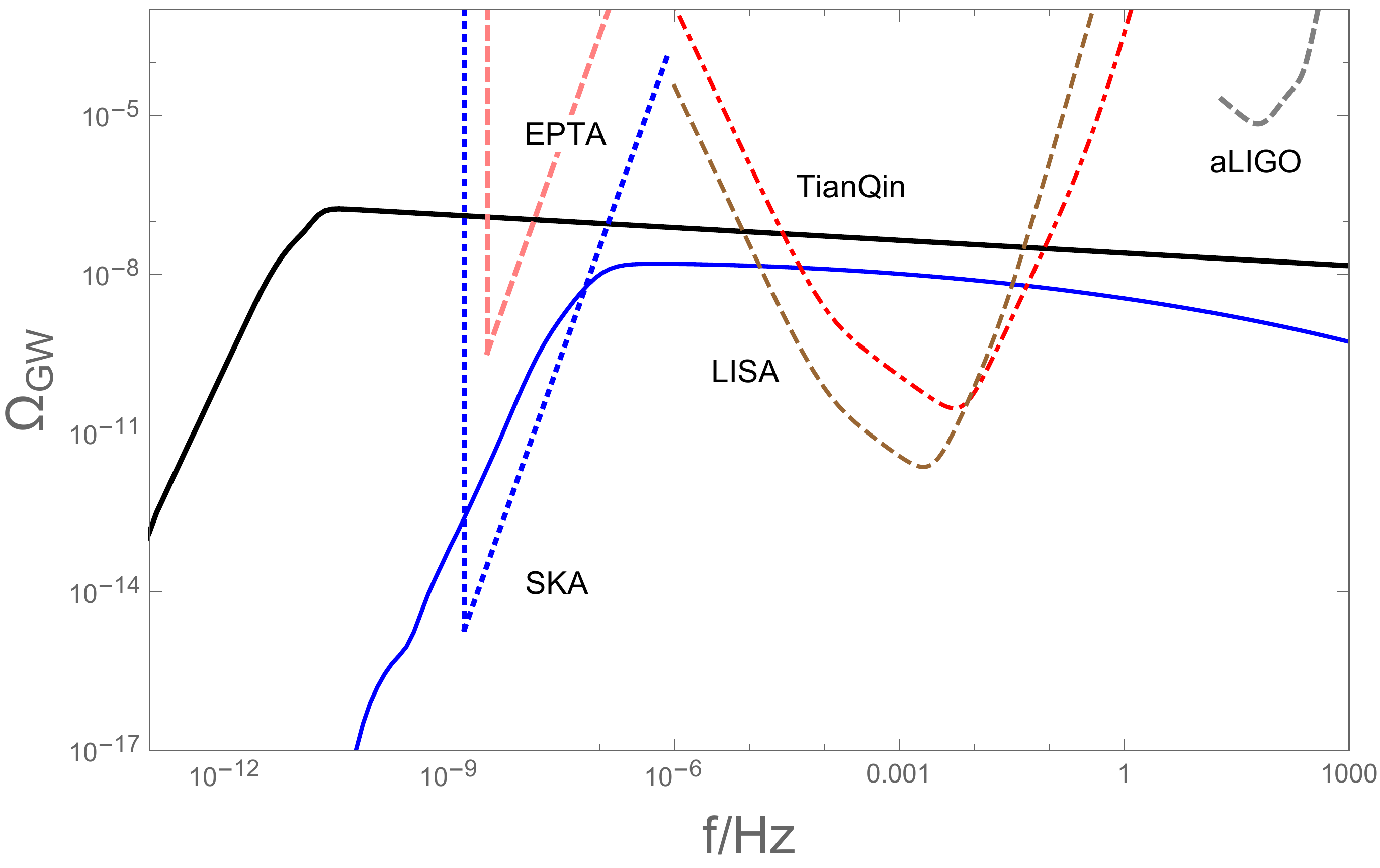}
    \caption{The secondary GW signal generated by density perturbations that produce PBH dark matter. The solid black line shows induced GWs from the piecewise parametrization constrained by PBH dark matter.
We also show induced GWs from a inflationary model \cite{Gong:2017qlj} by the solid blue line.
The sensitivity curves from different observations are also shown \cite{Sathyaprakash:2009xs,Moore:2014lga,Kuroda:2015owv}.
The pink dashed curve denotes the EPTA limit, the blue dotted curve denotes the SKA  limit, the red dot-dashed curve in the middle denotes the TianQin limit,
the brown dashed curve shows the LISA limit,
and the gray dashed curve denotes the aLIGO limit.
}
    \label{fig:results}
\end{figure}

Due to the uncertainties in the value of $\delta_c$ and the effect of non spherical collapse, the upper limit on the power spectrum
by the non detection of PBH dark matter can be much different \cite{Akrami:2016vrq,Sato-Polito:2019hws}. However, the method discussed
here can be easily applied to those cases. Using the power spectrum \eqref{pzeq3} and the method of calculating induced GWs presented in the previous section,
we obtain the energy density of secondary GWs and the result is shown in Fig. \ref{fig:results}. In Fig. \ref{fig:results},
we also plot the sensitivity curves for the ground based detector advanced Laser Interferometer Gravitational-Wave Observatory (aLIGO) \cite{Harry:2010zz,TheLIGOScientific:2014jea}, future space based GW detectors LISA \cite{Danzmann:1997hm,Audley:2017drz} and TianQin \cite{Luo:2015ght},
and PTA \cite{Kramer:2013kea,Hobbs:2009yy,McLaughlin:2013ira,Hobbs:2013aka}
including the European PTA (EPTA) and SKA \cite{Moore:2014lga}. It is obvious that the secondary GWs can be detected by EPTA, SKA, LISA and TianQin
although there is no detection of PBH dark matter.
In other words, the observation of induced GWs puts stronger constraint on the primordial curvature perturbation at small scales.
Since the current PTA observations don't find stochastic GWs yet, so the upper limit \eqref{pzeq3} for $k\gtrsim 10^4\t{ Mpc}^{-1}$ is overestimated. Using the power-law power spectrum \eqref{pzeq3},
we calculate the $\mu$ distortion \cite{Chluba:2015bqa,Nakama:2017ohe}
\begin{equation}
\label{mueq1}
\mu_{\text{ac}}\approx \int_{k_{\text{min}}}^\infty \frac{dk}{k}\mathcal{P}_\zeta(k)W_\mu(k),
\end{equation}
where
\begin{equation}
\label{mueq2}
W_\mu(k)=2.8A^2\left[\exp\left(-\frac{[\hat{k}/1360]^2}{1+[\hat{k}/260]^{0.3}+\hat{k}/340}\right)
-\exp\left(-\left[\frac{\hat{k}}{32}\right]^2\right)\right],
\end{equation}
$k_{\text{min}}\approx 1~$Mpc$^{-1}$, $A\approx 0.9$ and $\hat{k}=k/[1\ \text{Mpc}^{-1}]$, and we get $\mu_{\text{ac}}=0.03$. Again this result
shows that the upper limit \eqref{pzeq3} for $k\gtrsim 10^4\t{ Mpc}^{-1}$ is too large.
In other words, the upper limit set by PTA and $\mu$ distortion is more stringent.

For the power-law power spectrum, if there is no detection of induced GWs by LISA, then the constraint is
\begin{equation}
\label{pzeq4}
\mathcal{P}_\zeta\leq 3.9\times 10^{-4}\sbra{\frac{k}{1.8\times 10^{12} \text{ Mpc}^{-1}}}^{0.96-1}.
\end{equation}
If we choose $\delta_c=0.42$, plugging the constraint \eqref{pzeq4} into Eqs. \eqref{pbheq4} and \eqref{3}, we get $f_{\text{PBH}}<10^{-400}$.
This means if LISA does not observe induced GWs, then the contribution from PBHs with the mass around $10^{-14}M_{\odot}$ to dark matter is negligible.
In Fig. \ref{fig:results}, we also show the secondary GWs produced by
the inflationary model with the polynomial potential \cite{Gong:2017qlj}.
For convenience, we call the model as D-G model.
From Fig. \ref{fig:results}, we find that the D-G model can be tested by SKA, LISA and TianQin in the future.

\section{Conclusion}

In the case that the production of secondary GWs starts long
before the horizon reentry,
there was an analytical formula for the time integral of the source $I_{\text{RD}}(u,v,\eta)$.
For secondary GWs produced only after the horizon reentry,
we derive similar analytical formula for $I_{\text{RD}}(u,v,\eta)$
by splitting $I_{\text{RD}}(u,v,\eta)$ into the combinations of two oscillations $\sin(k\eta)$ and $\cos(k\eta)$.
With this analytical formula, it is easy to obtain the $1/\eta^2$ behavior of the power spectrum of induced GWs
and hence it helps to understand why induced GWs evolve as radiation at late time.
For nearly
scale invariant primordial curvature perturbations,
we find that the GWs produced before the horizon reentry contribute about 5\%
to the total energy density of induced GWs because all the modes accounts for the production. For the power spectrum of curvature perturbations with a sharp peak
which is parameterized as a delta function, there exists resonant amplification
because only one single mode contributes to the integration, and
the production of GWs before the horizon reentry becomes dominant at large scales.
Since the amplitude of induces GWs is proportional to the square of the peak
amplitude of the power spectrum of curvature perturbations, $A_\zeta^2$,
the upper bound on $A_\zeta$ from the observational constraints on PBH dark matter
can be used to discuss the possible detection of induced GWs.

Using the piecewise power-law parametrization for the power spectrum of primordial curvature perturbations and the observational constraints on PBH dark matter,
the best fit upper bound on primordial curvature perturbations was obtained.
We find that at small scales $k\gtrsim 10^4$ Mpc$^{-1}$, the upper limit on the power spectrum is $\mathcal{P}_\zeta\lesssim 0.05$.
However, this upper limit gives large stochastic GW background which is inconsistent with the observations of EPTA and the $\mu$ distortion caused by this upper limit is also too large.
This means that in nanohertz bands, PTA observations set more stringent upper bound.
On the other hand, if the power spectrum peaks at some particular small scales,
it evades the constraint by EPTA.
Therefore, the detection of induced GWs in the future puts more stringent constraint on the power spectrum and the abundance of PBH dark matter.
The non-detection of induced GWs by LISA constrains
the power spectrum in the LISA band to be $\mathcal{P}_\zeta\lesssim 4\times 10^{-4}$, so the contribution from PBHs with the mass around $10^{-14}M_{\odot}$ to dark matter is negligible
if induced GWs are not observed by LISA in the future.

\section{acknowledgments}
This research was supported in part by the National Natural Science
Foundation of China under Grant No. 11875136 and
the Major Program of the National Natural Science Foundation of China under Grant No. 11690021. The authors would like to thank Davide Racco for pointing out their
work which derived similar formulae as ours in Section 2.

\bibliographystyle{JHEP}
\bibliography{../../book/cosmologyref}

\providecommand{\href}[2]{#2}\begingroup\raggedright\endgroup

\end{document}